\documentclass[prl,showpacs,twocolumn,amsfonts]{revtex4}

\usepackage{graphicx}
\usepackage{bm}

\begin{document}

\title{Stretching weakly bending filaments with spontaneous curvature in two dimensions}
\author{Panayotis Benetatos and Eugene M. Terentjev}
\affiliation{Cavendish Laboratory, University of Cambridge, J. J.
  Thomson Avenue, Cambridge, CB3 0HE, United Kingdom }

\date{\today}

\begin{abstract}
Some important biomolecules (for instance, bacterial FtsZ and eukaryotic DNA)
are known to posses spontaneous (intrinsic) curvature. Using a simple
extension of the wormlike chain model, we study the response of a weakly
bending filament in two dimensions to a pulling force applied at its
ends (a configuration common in classical in-vitro experiments and
relevant to several 
in-vivo cell cases). The spontaneous curvature of such a chain or filament 
can in general be arc-length dependent and we study a case of 
sinusoidal variation, from which an arbitrary case can be reconstructed via
Fourier transformation. We obtain analytic results for the
force-extension relationship and the width of transverse fluctuations. We show that
spontaneous-curvature undulations can affect the force-extension behavior
even in relatively flexible filaments with a persistence length smaller than the contour length.

\end{abstract}
\pacs{36.20.Ey,87.15.ad,82.37.Rs}
\maketitle

\section{I. Introduction}

Semiflexible polymers are fluctuating macromolecules with a bending
stiffness intermediate between that of a random coil (Gaussian chain)
and a rigid rod. They have been the subject of intense experimental
and theoretical study in recent years because many important
biomolecules fall in this category. The elastic behavior of the structural elements of
the cytoskeleton (F-actin, microtubules, intermediate filaments) as
well as that of
DNA strongly depend on their bending stiffness \cite{JH,PhN}.

A widely used minimal model of semiflexible polymers is
the wormlike chain model (WLC) which treats the filament as a
locally inextensible one-dimensional fluctuating curve with bending
stiffness \cite{KP,Saito}. This model involves only two length scales, namely, the
contour length $L$ and the persistence length $L_p$, which is the
correlation length of the polymer directedness related to the chain
bending modulus $\kappa = \frac{1}{2} L_p k_BT(d-1)$, where $d$ is the
dimensionality of the embedding space. In the limit $L / L_p \gg 1$ the 
semiflexible chain transforms into a Gaussian coil and the question of
excluded-volume interaction may become relevant; in the opposite limit 
$L/L_p \ll 1$ the chain approaches the limit of rigid rod, in which case 
the additional modulus penalizing the local stretching/compression may 
come into play \cite{Blundell}. Despite its
simplicity, the WLC model has been successfully used to describe
some important experimental results such as the strong stretching of
ds-DNA \cite{Bustamante}. However, the advent of single molecule
experiments has revealed many occasions where the complexity of the
microscopic architecture of the filament appears to affect its
macroscopic behavior in a way that cannot be accounted for by the classical
WLC model. 

An important property of some biomolecules which goes beyond the basic
WLC model is spontaneous local curvature. 
FtsZ is a protein of the prokaryotic cytoskeleton which plays an
essential role in cell division as it assembles into the contractile
ring (Z-ring) that divides the bacterium \cite{Ghosh}. It can be
visualized on a mica surface using atomic-force microscopy (AFM). An
important property of many FtsZ protofilaments is the spontaneous
curvature of their relaxed state. Their role in force generation
strongly depends on the determination of their persistence length
\cite{Erickson}. Assuming the relaxed state of FtsZ protofilaments to
be straight, Dajkovic {\it et al.} \cite{Dajkovic} estimated $L_p
\approx 200\; {\rm nm}$ which is much too small. On the other hand, 
Horger {\it et al.} \cite{Horger}
taking into account the finite spontaneous curvature measured by AFM
(average radius of curvature $\approx 100\; {\rm nm}$), determined a
persistence length $L_p \approx 4\; {\rm {\mu}m}$ which is close to the 
value calculated from models of the flexural rigidity.

The existence of stable bent configurations of DNA has been known for
many years and they play important roles in  many basic
genetic processes \cite{Ohyama}. Spontaneous curvature in the local
structure of DNA is sequence-dependent \cite{Han_PNAS}. As in the case of
FtsZ, the presence (or absence) of local spontaneous curvature in a
DNA filament is crucial in determining the persistence length and its
relation to the bending rigidity. In a recent experiment, AFM imaging
of two-dimensional human DNA chains and comparison to measurements on
both intrinsically straight and uncorrelated HCV DNA has demonstrated
their spontaneous curvature \cite{Moukhtar}.

The theoretical study of small (Gaussian) thermal fluctuations of 
three-dimensional filaments
with spontaneous curvature and torsion has yielded analytical
expressions for the orientational correlation functions and the
persistence length \cite{Panyukov_Rabin}. An auxiliary field theory of
three-dimensional polymers with intrinsic curvature has obtained
the orientational correlations, mean square separation and
the force-extension relation of such chains in the $L / L_p \gg 1$ limit, 
showing that helical configurations arise due to the excluded-volume 
interaction \cite{Craig}. Real filaments with spontaneous curvature in
three dimensions also possess a finite twist rigidity which has been
shown to affect the end-to-end distribution \cite{KR_twist} and the
response to a stretching force giving rise to
instabilities \cite{KR_PRL,Netz_EPL}.
The zero-temperature conformations of two-dimensional filaments with constant
spontaneous curvature under an applied force for various boundary
conditions, as well as the linear
response at finite temperature to an applied force, have been 
recently investigated in \cite{Zhou}.

In this paper, we study the response of a two-dimensional WLC 
with spontaneous curvature to a stretching force applied at the ends. Although
in general one of course expects the stretched chain to explore fluctuations
in both perpendicular directions, the two-dimensional limit is a meaningful 
simplifying approximation that permits us to carry out a complete analytical
study, while retaining some of the relevant qualitative features of the general
problem. The three-dimensional problem is more complicated in two
nontrivial aspects. Curvature in $3d$ is always positive definite
whereas in $2d$ we can define the signed curvature. In addition, the presence of
twist rigidity complicates matters even further \cite{KR_twist,KR_PRL,Netz_EPL} . As we explain in Sec. VII,
however, the main
conclusion of this work which is the softening of the stretching
response due to a spatially varying spontaneous curvature is expected to carry through
in $3d$. Besides, many imaging experimental investigations
\cite{Horger,Moukhtar} are carried out on molecules confined to a
surface. A stretching experiment on DNA electrostatically confined to
the surface of a fluid membrane is presented in \cite{Radler}.
The elasticity of semiflexible polymers without spontaneous
curvature in two dimensions has been studied analytically in
\cite{Kondev}. The spontaneous curvature can in
general be inhomogeneous, dependent on the location along the arc length
$s$ along the chain. Any such variation can be decomposed into Fourier 
harmonics and so we study a case of local curvature varying in 
an arbitrary sinusoidal fashion along the chain length. We
assume that the filament is weakly bending. This approximation can be
realized in the case of filaments with small spontaneous curvature and
small thermal fluctuations or
in the case of strong stretching forces which have flattened out the
initially large spontaneous curvature and/or the large thermal undulations. This is the simplest possible
extension of the WLC model to include spontaneous curvature
analytically. A formally similar approach has recently been employed
in \cite{Zhou2} to study uncorrelated quenched disorder in the
spontaneous curvature of stretched two-dimensional filaments.

The paper is organized as follows. In Sec. II we introduce our model
and derive the Gaussian energy functional which characterizes the
stretched polymer. In the absence of thermal fluctuations, the polymer
assumes the ground-state 
conformation which minimizes the elastic energy for given
boundary conditions; finding this conformation
is the subject of Sec. III. In Sec. IV we
show that thermal fluctuations
decouple from the athermal mechanical response and are independent of the
spontaneous curvature. In Sec. V we calculate the force-extension
relationship and identify the various force scales in the problem,
while in Sec. VI we investigate the response of the shape of transverse
fluctuations on the stretching force. We conclude and discuss a
possible extension of this work in Sec. VII.

\section{II. Model} 
\begin{figure}
\includegraphics[width=0.3\textwidth]{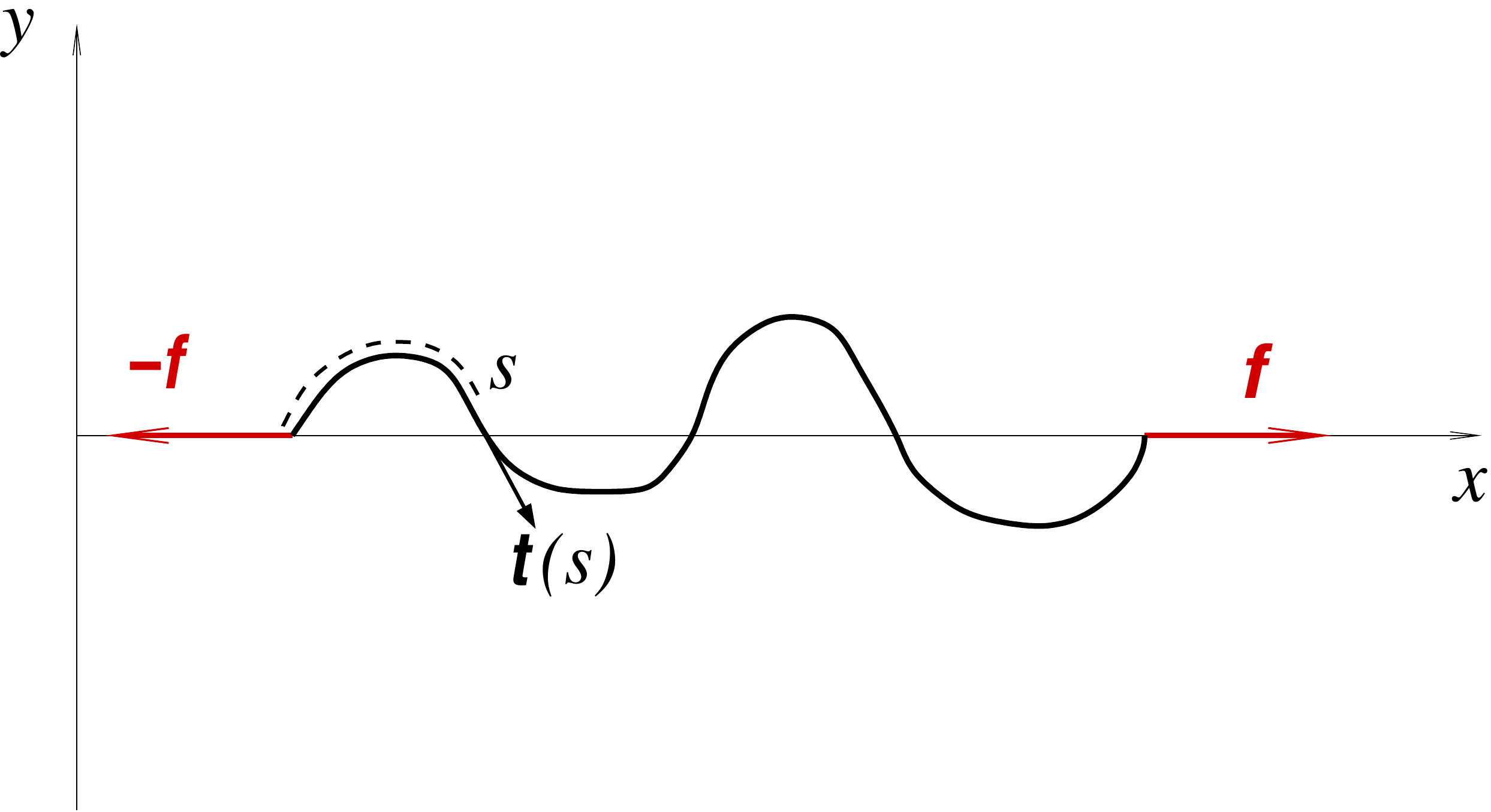}
\caption{Schematic diagram of a filament with spatially varying
  spontaneous curvature showing the notation used.\label{scs_fig1}}
\end{figure}

We treat a semiflexible polymer with spontaneous
(intrinsic) curvature by simply extending the classical WLC
model. The filament is described as a plane curve, ${\bf r}(s)$, 
where $s \in [0,L]$ is the arc-length along its contour, subject
to a local inextensibility constraint so that the total contour 
length $L$ is constant. The constraint is mathematically expressed 
by the condition:
\begin{equation}
\label{inext_constraint}
\Big|\frac{\partial {\bf r}(s)}{\partial s}\Big|=1\;.
\end{equation}
The effective free-energy functional (``Hamiltonian'') of the polymer
depends only on its curvature (bending),
\begin{equation}
\label{H_0}
{\cal H}_0[{\bf r}(s)]=\frac{\kappa}{2}\int_0^L ds\, \Big[\Big|\frac{\partial {\bf
    t}(s)}{\partial s}\Big|-c(s)\Big]^2\;,
\end{equation}
where ${\bf t}(s)={\partial {\bf r}(s)}/{\partial s}$ is the
tangent vector and  $c(s)$ is the spontaneous curvature at the
position $s$, cf. Fig.\ref{scs_fig1}. This bending energy
penalizes conformations with a local curvature if it deviates away from 
$c(s)$.

The interaction with a stretching force ${\bf f}$ which pulls at the polymer
end-points is expressed by adding to ${\cal H}_0$ the work function
\begin{eqnarray}
\label{H_I}
{\cal W}[{\bf r}(s)]=-{\bf f}\cdot[{\bf r}(L)-{\bf r}(0)]=-\int_0^L ds\, {\bf f}\cdot {\bf t}(s)\;.
\end{eqnarray}

The stretching response of a filament described by Eqs. (\ref{H_0}) and (\ref{H_I})
is in general analytically intractable for two reasons. One is the well-known
problem associated with the local inextensibility constraint, the
other is the non-analyticity introduced by the definition of
curvature as the modulus of $d {\bf t}(s)/ds$. However, in the present case of
weakly-bending filaments in two dimensions the problem simplifies
drastically and becomes linear. The coordinate system that we use is
shown in Fig. \ref{scs_fig1}.
To describe a weakly-bending filament, it is
convenient to use the parametrization ${\bf r}(s)=[s-X(s),\; y(s)]$ which
implies ${\bf t}(s)=[1-\partial_s X,\; t_{\perp}(s)]$. The inextensibility
condition can be written as $t_{\perp}^2+ ( 1-\partial_s X)^2=1$ which, in the
weakly bending approximation $|t_{\perp}(s)|\ll 1$, yields
$\partial_s X=\frac{1}{2}t_{\perp}^2+{\cal O}(t_{\perp}^4)$ and therefore
$\partial_s^2X=t_{\perp}\partial_s t_{\perp}+
\frac{1}{2}t_{\perp}^3\partial_s t_{\perp}+ {\cal O}(t_{\perp}^5)$. In
this approximation, the principal normal to the filament can be
considered perpendicular to its backbone ($x$-direction) and $\partial{\bf
  t}/\partial s\approx[0,\; \partial_s t_{\perp}]$. Introducing the
signed curvature, ${\tilde c}(s)$, we can express the energy of a
weakly-bending stretched filament as
\begin{eqnarray}
\label{H_wb}
{\cal H}[y(s)]&= &\frac{\kappa}{2}\int_0^L ds \Big[\Big(\frac{\partial^2
  y(s)}{\partial s^2}\Big)-{\tilde c}(s)\Big]^2\nonumber\\
& &+\frac{1}{2}f\int_0^L
  ds \Big(\frac{\partial
  y(s)}{\partial s}\Big)^2 -fL\;.
\end{eqnarray}
${\tilde c}(s)$ is positive if the filament bends convexly and
  negative if it bends concavely.

\section{III. Ground state}

In the absence of thermal fluctuations, the response of the filament
to the stretching force is determined
by minimizing the functional ${\cal H}[y(s)]$ for appropriate
boundary conditions. In this paper, we assume hinged-hinged boundary
conditions  which can be realized by attaching beads at the end-points
that can exert forces but not moments \cite{upenn_eng}. As we show
below, in the case of  strong stretching, the specific boundary
conditions may become irrelevant.  Formally, they are expressed by
the following relations:
\begin{eqnarray}
\label{bcs}
&&y(0)=y(L)=0\nonumber \\
&&y''(0)={\tilde c}(0),\; y''(L)={\tilde c}(L)\;.
\end{eqnarray}
(The primes denote derivatives with respect to $s$.) The minimizing
configuration, $y_a(s)$,  obeys an Euler-Lagrange equation which
can be obtained by adding an infinitesimal variation,
$y(s)=y_a(s)+\epsilon \eta(s)$, where $\epsilon$ is infinitesimal and
$\eta(s)$ is an arbitrary function of $s$ satisfying
$\eta(0)=\eta(L)=0$ and $\eta''(0)=\eta''(L)=0$ and requiring the
derivative of ${\cal H}({\epsilon})$ as a function of ${\epsilon}$ to vanish. The athermal
configuration is the solution to the equation
\begin{eqnarray}
\label{E-L}
-f y_a''(s)+\kappa y_a''''(s)-\kappa {\tilde c}''(s)=0
\end{eqnarray}
with the boundary conditions given by Eq. (\ref{bcs}).

We first consider the case of constant curvature: the value ${\tilde
  c}(s)=c_0$ is maintained along the chain. The last term in Eq. (\ref{E-L}) 
vanishes and the athermal configuration $y_a(s)$ is determined by
hyperbolic functions with the characteristic length scale $l_f\equiv \sqrt{\kappa/f}$. 
Implementing the boundary conditions, we obtain \begin{eqnarray}
\label{const_c_ath}
&&y_a(s)=\frac{c_0 \kappa}{f
  \sinh\big(\frac{L}{l_f}\big)}\Big[\sinh\big(
  \frac{s}{l_f}\big)+\sinh\big(
  \frac{(-s+L)}{l_f}\big)\nonumber\\
&&-\sinh\big(\frac{L}{l_f}\big)\Big]\;.
\end{eqnarray}
As expected, in the vanishing-force limit $f \rightarrow 0$
the above expression yields a parabola, $y_a(s)=\frac{1}{2}c_0s(s-L)$,
which is identical, to leading order in the weakly bending
approximation, to a circular arc of curvature $c_0$. For
$l_f\ll L$, and $l_f\ll s \ll L-l_f$, we get a uniform displacement,
$y_a(s)=-c_0{\kappa}/f$, which implies that the actual response comes
from the segments of arc-length $l_f$ at the boundaries.

For a filament with arbitrary variable spontaneous curvature, one can expand it
in Fourier modes:
\begin{equation}
\label{Fourier}
{\tilde c}(s)=\sum_n c_n \cos(qs+\phi_n),\;q\equiv \frac{n\pi}{L},\;n \in \mathbb{N}\;.
\end{equation}
Since Eq. (\ref{E-L}) is linear, its solution for the general case
will be the superposition of the modes which correspond to the
individual harmonics. Taking a particular mode, ${\tilde c}(s)=c_n
\cos(qs+\phi)$, the ground-state stretched configuration becomes:
\begin{widetext}
\begin{eqnarray}
\label{ath_cos_phi}
y_a(s)=\frac{{\kappa}c_n}{(f+q^2{\kappa})\sinh(\frac{L}{l_f})}
\Big[2\cos({\phi})\sinh\big(\frac{1}{2}\frac{L}{l_f}\big)
\cosh\big(\frac{1}{2}\frac{(L-2s)}{l_f}\big)-
\cos(qs+{\phi})\sinh(\frac{L}{l_f})\Big].
\end{eqnarray}
\end{widetext}
Since Eq. (\ref{E-L}) is linear, a chain with spontaneous curvature
containing many harmonic modes yields a response ($y_a(s)$) which is
the superposition of the responses for the individual modes. The above
expression becomes more
transparent in the case where ${\tilde c}(s)=c_0\sin(qs)$,
\begin{eqnarray}
\label{ath_sin}
y_a(s)=-\frac{c_0{\kappa}\sin(qs)}{f+q^2{\kappa}}\;.
\end{eqnarray}
In the strong stretching regime, $l_f\ll L$, any mode yields a uniform
response for  $l_f\ll s \ll L-l_f$, 
\begin{eqnarray}
\label{ath_strong_stretch}
y_a(s)=-\frac{c_n{\kappa}\cos(qs+\phi_n)}{f+q^2{\kappa}}\;,
\end{eqnarray}
consisting of a simple reduction in the amplitude. We see that the
length scale $l_f$ defines the distance along the chain over which the
boundary effects persist (boundary layer).

\section{IV. Response at finite temperature}
Thermal fluctuations are added to the athermal ground-state response,
$y(s)=y_a(s)+u(s)$, and contribute according to a Boltzmann weight
associated with the energy given in Eq. (\ref{H_wb}). As we show
below, in the
linear theory of the weakly-bending approximation, the two
Gaussian contributions (thermal and athermal) decouple. 
The elastic energy of thermal fluctuations then takes the form,
\begin{eqnarray}
\label{decoupling}
&&{\cal H}[u(s)] = \frac{\kappa}{2}\int_0^L ds\, (u''(s))^2+\frac{f}{2}
\int_0^L ds\, (u'(s))^2\nonumber\\
&&+{\kappa}\int_0^L ds\,u''(s)\big(y_a''(s)-{\tilde
  c}(s)\big)+f\int_0^Lds\,y_a'(s)u'(s)\nonumber\\
&&+\,const\;.
\end{eqnarray}
Using the Euler-Lagrange equation (\ref{E-L}), and the boundary
conditions (\ref{bcs}), one can easily show that the
second line of the above equation vanishes. Therefore $\langle
u(s)\rangle =\langle u'(s)\rangle=\langle u''(s)\rangle=0$, where $\langle ...\rangle$ denotes
average with the weight $\exp(-{\beta}{\cal H}[u(s)])$,
${\beta}\equiv 1/k_B T$.

We calculate $\langle u^2(s)\rangle$ and $\langle
(u'(s))^2\rangle$ by decomposing $u(s)$ into Fourier harmonics which respect
the prescribed boundary conditions: $u(s)=\sum_n a_n \sin(n{\pi}s/L)$,
$n \in \mathbb{N}$. This gives:
\begin{eqnarray}
\label{u'_sq}
\langle (u'(s))^2\rangle=\frac{2k_B T}{L}\sum_{n=1}^{\infty}
\frac{{\cos}^2(\frac{n\pi}{L}s)}{[\kappa(\frac{n\pi}{L})^2+f]}
\end{eqnarray}
and
\begin{eqnarray}
\label{u_sq}
\langle (u(s))^2\rangle=2k_B T L\sum_{n=1}^{\infty}\frac{{\sin}^2
(\frac{n\pi}{L}s)}{[\kappa(\frac{n\pi}{L})^2+f](n\pi)^2}\;.
\end{eqnarray}
Note that the spontaneous curvature $\tilde{c}$ does not affect these
thermal fluctuations. The decoupling of thermal and athermal curvature 
undulations and the results
that we have obtained so far allow us to calculate various
experimentally relevant observables
such as force-extension curves and the average width of transverse fluctuations.

\section{V. Force-extension relationship}
The average projected length of the filament in the
direction of the stretching force is given by
\begin{eqnarray}
\label{ave(x)}
&&\langle x(L) \rangle = L - \frac{1}{2}\int_0^L ds \langle t_{\perp}^2(s)
\rangle \nonumber\\
&&=L-\frac{1}{2}\int_0^L ds  (y'_a(s))^2-\frac{1}{2}\int_0^L ds \langle (u'(s))^2\rangle,
\end{eqnarray}
where we have used the weakly bending approximation.
There are four different force scales in the general problem,
\begin{eqnarray}
\label{force_scales}
f_{cr}\equiv \frac{\kappa}{L_p^2},&\;\;\;&f_c\equiv \frac{\kappa}{c_0^{-2}}\nonumber\\
f_q\equiv \frac{\kappa}{q^{-2}},&\;\;\;&f_L\equiv \frac{\kappa}{L^2}
\end{eqnarray}
associated with the corresponding length scales: the persistence
length $L_p=2\kappa/k_BT$, the radius of spontaneous curvature
$1/c_0$, the wavelength of modulation of the spontaneous curvature
$2{\pi}/q$, and the contour length $L$. The force scale $f_L$ is
related to finite-size effects in the sense that for $f\gg f_L$, or
equivalently for $l_f\ll L$, the filament can be viewed as an
effective freely jointed chain consisting of ``pieces'' each 
of contour length $l_f$  \cite{Kierfeld,deGennes}. In principle, the length scales $1/c_0$ and $1/q$ are
independent (the former is the amplitude of the spontaneous curvature
whereas the latter is the wavelength). However, the case where $c_0 \gg q$ corresponds to very
steep undulations and goes beyond the weakly bending
approximation in the absence of a strong stretching force. The steepest undulations compatible with this
approximation have $c_0\approx q$. On the other hand, we can be within
the weakly bending approximation with $c_0 > q$ if the stretching
force is strong enough.

\begin{figure}
\includegraphics[width=0.4\textwidth]{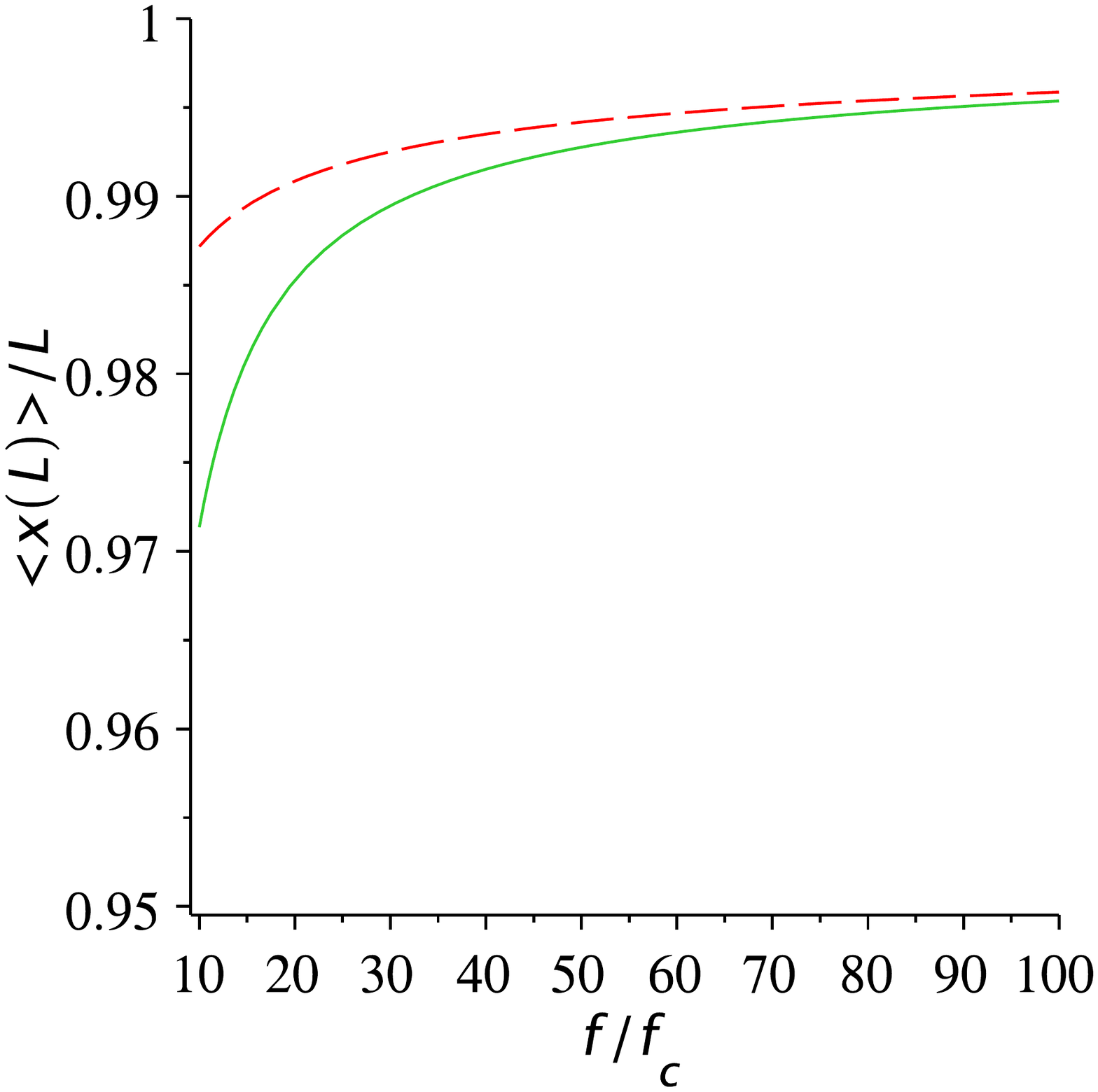}
\caption{Force-extension curve (solid line) for a filament with
  $L_p/L=1$ and $c_0L=12$. For comparison, we show the force-extension curve of the
  corresponding filament without spontaneous curvature (dashed line).
\label{force_ext0}}
\end{figure}

Let us first consider the case of constant spontaneous
curvature $\tilde{c}=c_0$. In the limit where $f\gg f_L$, the force-extension
relationship (\ref{ave(x)}) acquires the simple form,
\begin{eqnarray}
\label{f_ext1}
\frac{\langle x(L) \rangle}{L}=1-\frac{1}{4}\Big(\frac{L}{c_0^{-1}}\Big)
\Big(\frac{f_c}{f}\Big)^{3/2}-\frac{1}{2}\Big(\frac{f_{cr}}{f}\Big)^{1/2}.
\end{eqnarray}
The first term in the right-hand side (rhs) of Eq. (\ref{f_ext1}) is associated with
athermal stretching whereas the second term is associated with the
stretching of thermal fluctuations. The crossover force scale $\frac{1}{2}
({L}/{c_0^{-1}})({f_c^{3/2}}/{f_{cr}^{1/2}})\sim {\kappa} L_p
L c_0^4$ determines the relative weight of the two
contributions. In the strong stretching regime, the second term (which
does not depend on the spontaneous curvature) becomes dominant in
agreement with a similar result obtained recently in Ref. \cite{Zhou}. If
$L_p \approx L$ and the spontaneous curvature of the unstretched
filament is small, $c_0^{-1}\gtrsim L$, the athermal contribution is
almost negligible. It becomes significant, however, when the 
spontaneous curvature is large (the weakly bending
approximation still applies
to the stretched configuration of the filament). This is illustrated
in Fig. \ref{force_ext0}.

\begin{figure}
\includegraphics[width=0.4\textwidth]{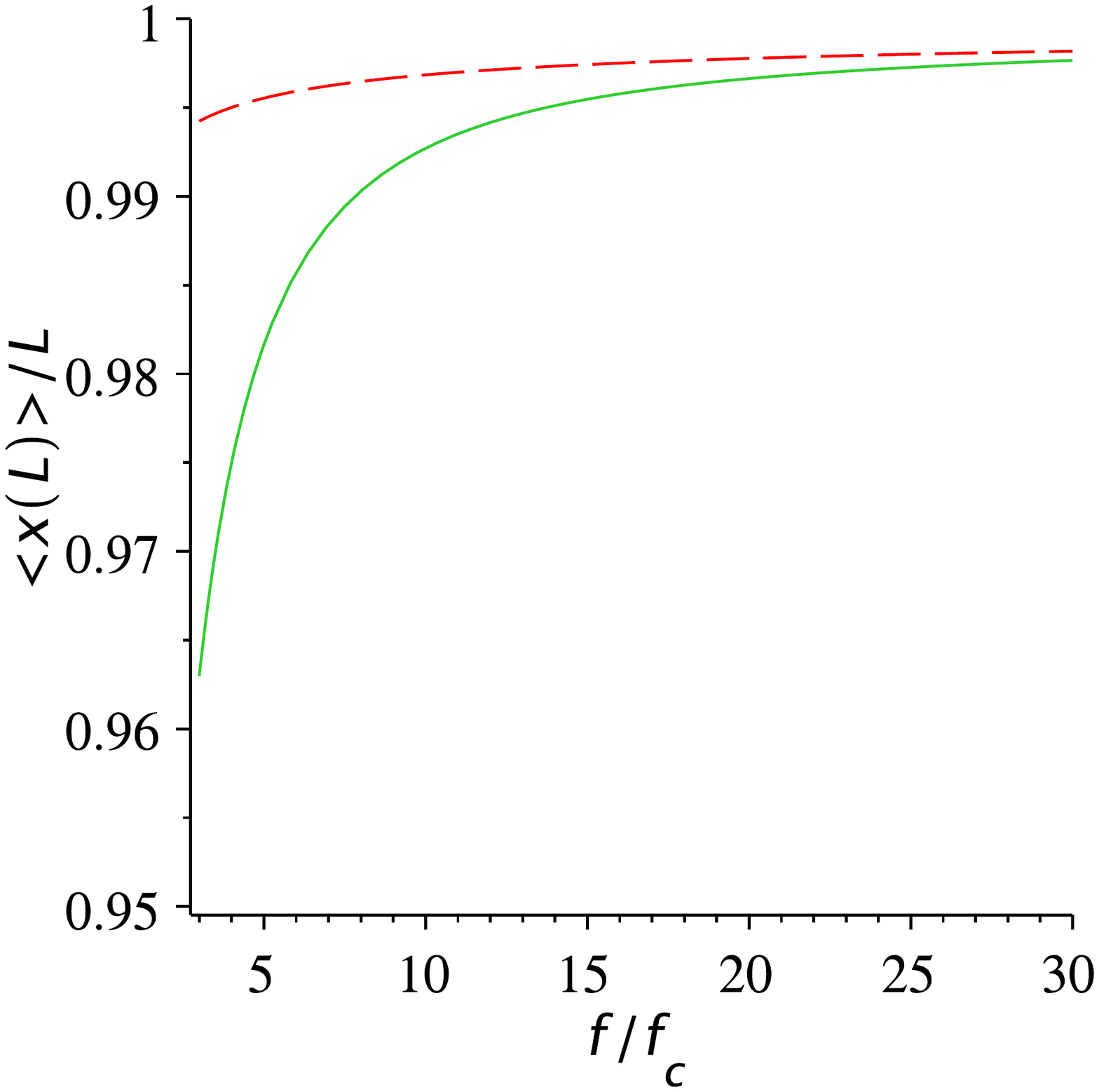}
\caption{Force-extension curve (solid line) for a filament with
  $L_p/L=0.05$ and $c_0L_p=q L_p =50$.  For comparison, we show the force-extension curve of the
  corresponding filament without spontaneous curvature (dashed line).
\label{force_extI}}
\end{figure}

We now consider the case of a filament with spontaneous
curvature which varies along its contour. If the local spontaneous curvature vanishes at the
end-points, we can expand it in a sine series. In this case, the 
force-extension relationship takes the form,
\begin{eqnarray}
\label{f_ext2}
&&\frac{\langle x(L)
  \rangle}{L}=1-\frac{1}{2}\frac{{\kappa}^2{\pi}^2}{L^2}\sum_{n=1}^{\infty}\frac{c_n^2 n^2}{[\kappa(\frac{n\pi}{L})^2+f]^2}\nonumber\\
&&-\frac{1}{2}\frac{k_B T}{L}\sum_{n=1}^{\infty}\frac{1}{[\kappa(\frac{n\pi}{L})^2+f]}\;.
\end{eqnarray}
We see that as the stretching force increases, it successively irons
out the larger length-scale (softer) modes of curvature (spontaneous or
thermally excited) \cite{Kabla}. If the spontaneous curvature has only
one mode and $f\gg f_L$, we obtain:
\begin{eqnarray}
\label{f_ext3}
\frac{\langle x(L) \rangle}{L}=1-\frac{1}{2}\frac{f_q f_c}{[f_q+f]^2}-
\frac{1}{2}\Big(\frac{f_{cr}}{f}\Big)^{1/2}.
\end{eqnarray}
This result holds irrespective of the value of the local spontaneous curvature at the end-points.
As expected, when the spontaneous-curvature undulations have been
ironed out by a strong force, the force-extension relationship is
determined by the residual thermal fluctuations. 
The same is true when the persistence
length is so small that thermal fluctuations wash out the spontaneous
curvature. However, when $L_p\gg c_0^{-1}\approx q^{-1}$, the spontaneous
curvature strongly affects the response to a stretching force. We
point out that for $f_L\ll f_{cr}\ll f \lesssim f_c \approx f_q$ the response
will be quite soft despite the nonlinear stretching of thermal
undulations due to $f_{cr}\ll f$ as illustrated in
Fig. \ref{force_extI}. If the spatially variable spontaneous curvature
contains many harmonic modes, our analysis still applies: Eqs. (\ref{ath_cos_phi}) and
(\ref{ave(x)}) yield a result similar to that of Eq. (\ref{f_ext2}) but
  with the first sum in the rhs replaced by a more complicated term.
\begin{figure}
\includegraphics[width=0.4\textwidth]{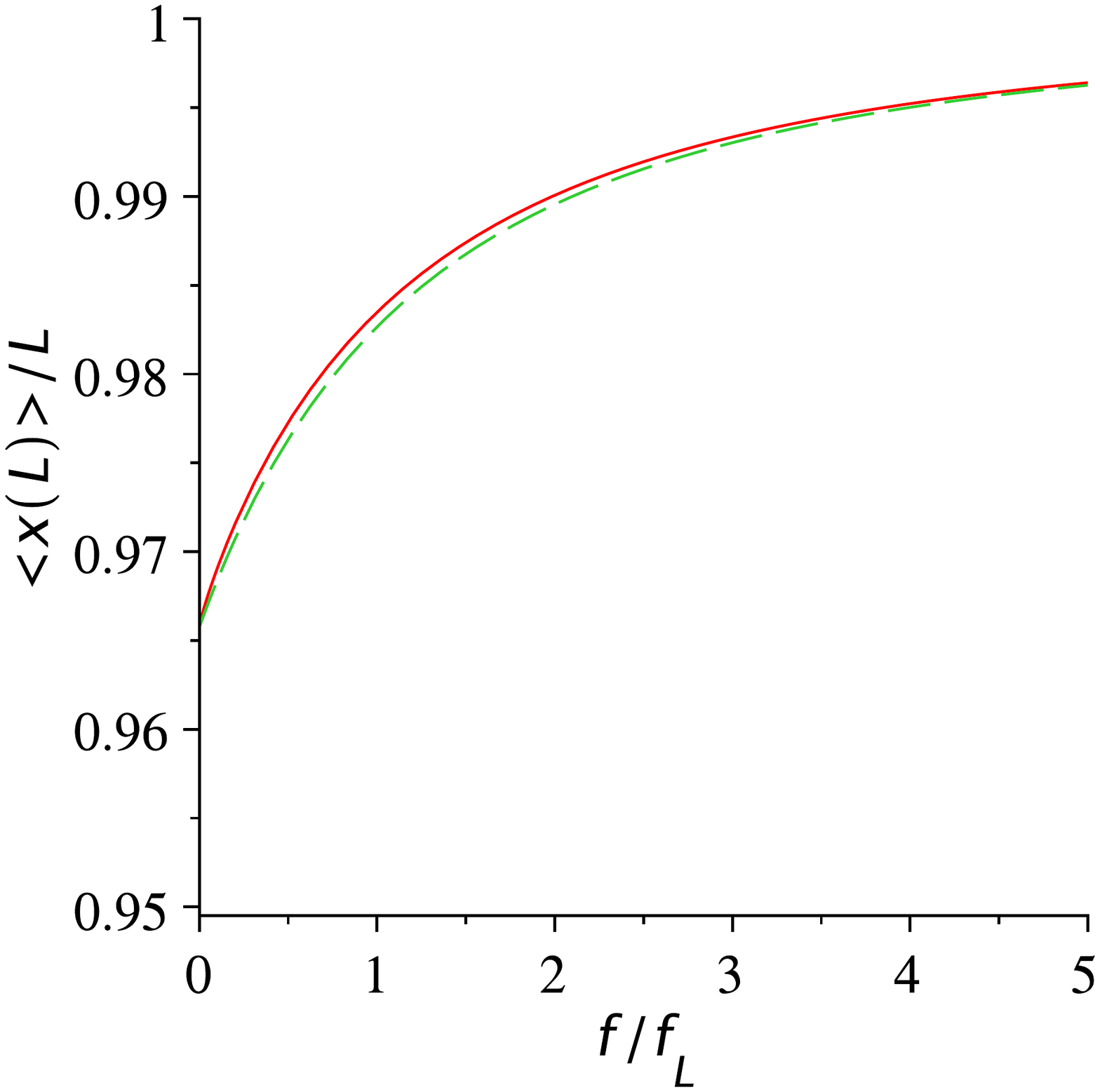}
\caption{Force-extension curve for a filament at $T=0$ with
  spontaneous curvature ${\tilde c}(s)=\frac{2.6}{L}\sin(\frac{10{\pi}}{L}s)$
  (solid line) and  ${\tilde c}(s)=\frac{2.6}{L}\cos(\frac{10{\pi}}{L}s)$
  (dashed line). 
\label{phase}}
\end{figure}
For  $f_L \gtrsim f$, the value of the local spontaneous  curvature at the end-points only slightly affects the response as
shown in Fig. \ref{phase}. 

We can gain insight into the effect of the spatially modulated
spontaneous curvature on the strong stretching response by
considering the effective freely jointed chain mentioned above. A
two-dimensional freely jointed chain consisting of $N$ ``pieces'' of
end-to-end distance $b$, has a force extension relationship,
\begin{eqnarray}
\label{FJC1}
\frac{\langle x(L)\rangle}{Nb}=1-\frac{k_BT}{2fb},
\end{eqnarray}
where $f\gg k_BT/b$ \cite{Kierfeld}. This equation can also be written
as 
\begin{eqnarray}
\label{FJC2}
\frac{\langle x(L)\rangle}{L}=\frac{b}{2l_f}-\frac{1}{2}\Big(\frac{f_{cr}}{f}\Big)^{1/2},
\end{eqnarray}
where the chain is viewed as a succession of $N$ ``pieces'' of
arc-length $2l_f$.
In the case of a WLC {\it without spontaneous curvature}, $b\equiv
2l_f$ \cite{Kierfeld}. In the case of a filament with spatially modulated
spontaneous curvature, however, the ``pieces'' will be wavy with $b <
2l_f$. Plugging the result of Eq. (\ref{ath_strong_stretch}) into the
athermal part of Eq. (\ref{ave(x)}), and assuming $q^{-1}\ll 2l_f$, we
obtain $b\equiv \langle x(2l_f)\rangle\approx 2l_f(1-f_qf_c/2[f_q+f]^2)$, thus
recovering Eq. (\ref{f_ext3}).

We note that in the case of spatially varying spontaneous curvature,
the force-extension response for $f\gg f_L$ comes mainly from ironing
out undulations in the bulk. Therefore we do not expect clamped
boundary conditions of small compliance (e.g., with the direction of
the clamped segments close to that of the pulling force), to significantly
affect the response. On the other hand, as shown in
Eq. (\ref{const_c_ath}), for a filament with constant spontaneous
curvature and large persistence length in the strong stretching regime
($f\gg f_L$), the response
comes mainly from the boundary layers and the results will depend on the
specific boundary conditions.

\section{VI. Transverse fluctuations}
Apart from the force-extension curve, the shape of transverse
fluctuations can be a useful diagnostic tool for the elasticity of a
semiflexible filament \cite{upenn_eng}. For a fluctuating filament
with spontaneous curvature ${\tilde c}(s)=c_0\sin(qs)$ ($q=n{\pi}/L$),
the average width of transverse fluctuations along the polymer contour
is given by
\begin{eqnarray}
\label{transv_fluc}
&&\frac{1}{L}\int_0^L ds \langle
(y(s))^2\rangle=\frac{{\kappa}f_c}{(f+f_q)^2}\nonumber\\
&&+\;L k_B T\frac{({\tilde f})^{3/2}-3\coth(\sqrt{{\tilde f}}){\tilde
    f}+3{\sqrt{{\tilde f}}}}{6f_L({\tilde f})^{5/2}}\;,
\end{eqnarray}
where the non-dimensional shorthand ${\tilde f}\equiv f/f_L$ is used. 
The first (athermal) term is related to the spontaneous
curvature and the second term comes from the thermal fuctuations. For
$f\gg f_L$, the latter can be approximated by $\frac{1}{6}L k_B
T/f$. 

The profile of transverse fluctuations has the general form:
\begin{eqnarray}
\label{transv_fluc_prof}
&&\frac{\langle
(y(s))^2\rangle}{L}=\frac{c_0^2
  L^2}{(f/f_L)^2}F(f/f_{L,q},s/L)\nonumber\\
&&+\;\frac{L}{L_p}G(f/f_L,s/L),
\end{eqnarray}
where $F$ involves $f_L$ or $f_q$ depending on whether the spontaneous
curvature is constant or oscillating. In the
absence of spontaneous curvature, the profile of transverse
fluctuations is determined by the function $G(f/f_L,s/L)\propto
\langle (u(s))^2\rangle$ given in Eq. (\ref{u_sq}) which is  bell-shaped and its height decreases rather quickly
with increasing force. This also applies when the stretching force is
so strong that it has ironed out the undulations due to the
spontaneous curvature. Interestingly, this behavior persists even for
stretching forces such that $f_L \ll f_{cr} \ll f \ll f_c \sim
f_q$ where, as we have shown, the force-extension relationship is
strongly affected by the spontaneous-curvature contribution. The
behavior becomes qualitatively different in the case of rather stiff
filaments ($L_p\gg L$) for $f\ll f_q\sim f_c$ and $f/f_q \gg
L/L_p$. The athermal term becomes dominant and it decreases very
slowly with increasing force. In addition, the profile is
essentially flat, characterized by the constant amplitude of the
sinusoidal undulations.  This type of behavior is expected
also in the case
where $f_c \gg f_q$ and $f \ll f_q$ such that $f_c/f_q^2 \gg L/(L_p
f)$.  In order to get $f_c \gg f_q$, we need a filament with very
steep undulations of the spontaneous curvature ($c_0 \gg q$). This
would take us out of validity of the weakly bending
approximation. Although we cannot make quantitative predictions for
this case, we expect the main qualitative features of the predicted
behavior to persist.

\section{VII. Conclusions and Outlook}
In the present work, we considered a straightforward extension of the
WLC model which incorporates spontaneous curvature to study the
response of a two-dimensional weakly bending filament to a pulling
force applied at its ends. The response of a filament with intermediate flexibility
($L / L_p \approx 1$) and constant spontaneous curvature is affected
appreciably only for large initial curvature, where the effective stretching
modulus of the chain softens. If the 
spontaneous curvature is inhomogeneous, characterized by undulations of a characteristic
wavelength, we find a range of forces where these undulations  soften
the response despite the nonlinear flattening out of thermal fluctuations. 
The profile of transverse fluctuations is
mostly dominated by the bell-shaped thermal contribution unless the
spontaneous curvature undulations are very steep. 

Although our
quantitative results are based on the two-dimensional weakly bending
assumptions, the qualitative features are expected to hold more
generally as they are related to the length and force scales of the
problem. As we have shown in Sec. V, the softening of the stretching
response for $f_L\ll f_{cr} \ll f \ll f_c \approx f_q$ can be
attributed to the fact that the ``pieces'' of the effective freely
jointed chain have an end-to-end distance smaller than the
corresponding contour length because of their spontaneous
curvature. One might expect a
similar hierarchy of length and force scales to persist in
the case of a three-dimensional stretched helical filament with the
pitch length and the radius playing roles analogous to $q^{-1}$ and
$c_0^{-1}$, respectively, thus yielding a similar softening. Checking
the validity of 
this expectation is an interesting direction for future work.

\section{Acknowledgments}

P.B. gratefully acknowledges support from
EPSRC via the University of Cambridge TCM Programme Grant.


\begin{thebibliography}{}

\bibitem{JH}
J. Howard, {\it Mechanics of Motor Proteins and the Cytoskeleton}, (Sinauer Associates, Sunderland, MA, 2001).

\bibitem{PhN}
P. Nelson, {\it Biological Physics; Energy, Information, Life}, (W. H. Freeman \& Co., New York, NY, 2004).

\bibitem{KP}
O. Kratky and G. Porod, Recl. Trav. Chim. Pays-Bas {\bf 68}, 1106 (1949).

\bibitem{Saito}
N. Sait\^o, K. Takahashi and Y. Yunoki, J. Phys. Soc. Jpn. {\bf 22}, 219 (1967).

\bibitem{Blundell}
J. R. Blundell and E. M. Terentjev, Macromolecules {\bf 42}, 5388 (2009).

\bibitem{Bustamante}
C. Bustamante, S. B. Smith, J. Liphardt, and D. Smith, Curr. Opin. Struct. Biol. {\bf 10}, 279 (2000), and references therein.

\bibitem{Ghosh}
B. Ghosh and A. Sain, Phys. Rev. Lett. {\bf 101}, 178101, (2008).

\bibitem{Erickson}
H. P. Erickson, Proc. Natl. Acad. Sci. U.S.A. {\bf 106}, 9238, (2009).

\bibitem{Dajkovic}
A. Dajkovic, G. Lan, S. X. Sun, D. Wirtz, and J. Lutkenhaus,
Curr. Biol. {\bf 18}, 235, (2008).

\bibitem{Horger}
I. Horger, E. Velasco, J. Mingorance, G. Rivas, P. Tarazona, and M.
Velez, Phys. Rev. E {\bf 77}, 011902 (2008). 

\bibitem{Ohyama}
T. Ohyama (Ed.), {\it DNA Conformation and Transcription}, (Springer, New
York, NY, 2005).

\bibitem{Han_PNAS}
W. Han, M. Dlakic, Y.-J. Zhu, S. M. Lindsay, and R. E. Harrington,
Proc. Natl. Acad. U.S.A. {\bf 94}, 10565 (1997)

\bibitem{Moukhtar}
J. Moukhtar, E. Fontaine, C. Faivre-Moskalenko, and
A. Arneodo, Phys. Rev. Lett. {\bf 98}, 178101 (2007).

\bibitem{Panyukov_Rabin}
S. Panyukov and Y. Rabin, Phys. Rev. Lett. {\bf 85}, 2404, (2000).

\bibitem{Craig}
A. Craig and E. M. Terentjev, Macromolecules {\bf 39}, 4557, (2006).

\bibitem{KR_twist}
D. A. Kessler and Y. Rabin, J. Chem. Phys. {\bf 118}, 897, (2003).

\bibitem{KR_PRL}
D. A. Kessler and Y. Rabin, Phys. Rev. Lett. {\bf 90}, 024301, (2003).

\bibitem{Netz_EPL}
H. Wada and R. R. Netz, Europhys. Lett. {\bf 77}, 68001, (2007).

\bibitem{Zhou}
Z. Zhou, Phys. Rev. E {\bf 76}, 061913 (2007).

\bibitem{Radler}
B. Maier, U. Seifert, and J. O. R\"adler, Europhys. Lett. {\bf 60}, 622, (2002).

\bibitem{Kondev} A. Prasad, Y. Hori and J. Kondev, Phys. Rev. E {\bf
  72}, 041918 (2005).

\bibitem{Zhou2} 
Z. Zhou and B. Jo\'os, Phys. Rev. E {\bf 80}, 061911 (2009). 

\bibitem{Kierfeld}
J. Kierfeld, O. Niamploy, V. Sa-yakanit, and R. Lipowsky, Eur. Phys. J.
E {\bf 14}, 17, (2004).

\bibitem{deGennes}
P. G. de Gennes, in {\it Polymer Liquid Crystals}, edited by A. Ciferri,
W. R. Kringbaum and R. B. Meyer (Academic Press, New York, 1982)
Chapter 6.

\bibitem{Kabla}
A. Kabla and L. Mahadevan, J. R. Soc., Interface {\bf 4}, 99, (2007).

\bibitem{upenn_eng}
P. K. Purohit, M. E. Arsenault, Y. Goldman and H. H. Bau,
Int. J. Non Linear Mech. {\bf 43}, 1056 (2008).

\end{thebibliography}
\end{document}